# Terahertz light-matter interaction beyond unity coupling strength

*Andreas Bayer, Marcel Pozimski, Simon Schambeck, Dieter Schuh, Rupert Huber, Dominique Bougeard, and Christoph Lange[*]*

Department of Physics, University of Regensburg, 93040 Regensburg, Germany

[*]christoph.lange@physik.uni-regensburg.de

**Achieving control over light-matter interaction in custom-tailored nanostructures is at the core of modern quantum electrodynamics [1-15]. In ultrastrongly coupled systems [5-15], excitation is repeatedly exchanged between a resonator and an electronic transition at a rate known as the vacuum Rabi frequency $\Omega_R$. For $\Omega_R$ approaching the resonance frequency $\omega_c$, novel quantum phenomena including squeezed states [16], Dicke superradiant phase transitions [17,18], the collapse of the Purcell effect [19], and a population of the ground state with virtual photon pairs [16,20] are predicted. Yet, the experimental realization of optical systems with $\Omega_R/\omega_c \geq 1$ has remained elusive. Here, we introduce a paradigm change in the design of light-matter coupling by treating the electronic and the photonic components of the system as an entity instead of optimizing them separately. Using the electronic excitation to not only boost the oscillator strength but furthermore tailor the shape of the vacuum mode,**

**we push $\Omega_R/\omega_c$ of cyclotron resonances ultrastrongly coupled to metamaterials far beyond unity. As one prominent illustration of the unfolding possibilities, we calculate a ground state population of 0.37 virtual photons for our best structure with $\Omega_R/\omega_c$ = 1.43, and suggest a realistic experimental scenario for measuring vacuum radiation by cutting-edge terahertz quantum detection [21,22].**

In the strong coupling regime of quantum electrodynamics (QED), where the vacuum Rabi frequency $\Omega_R$ exceeds the dissipation rates of the electronic excitation and the resonator, new eigenmodes called cavity polaritons emerge. This universal principle is found in a large variety of systems, ranging from atoms [1] to excitons in semiconductors [2,3], molecules [4], mid-IR plasmonic structures [5-9], circuit QED systems at GHz frequencies [10-12], and structures in the THz spectral range [13-15].

In ultrastrongly coupled structures, $\Omega_R$ becomes comparable to the resonance frequency $\omega_c$ itself, the rotating-wave approximation of light-matter interaction falters, and anti-resonant coupling terms describing the simultaneous creation of correlated light and matter excitations become relevant [16,19,20]. Most prominently, the ground state is theorized to be a modified squeezed quantum vacuum with a finite population of correlated virtual photon pairs [16,20]. For sufficiently large values of the relative coupling strength $\Omega_R/\omega_c \gtrsim 1$, sub-cycle switching of $\Omega_R$ [6,9] may release these photons [16,20,23] in analogy to Unruh-Hawking radiation emerging at the event horizon of black holes [24]. These spectacular perspectives have fuelled the quest of the QED community for ever greater relative coupling strengths, ultimately aiming for $\Omega_R/\omega_c$ beyond unity.

The key strategy for boosting $\Omega_R/\omega_c$ comprises increasing the dipole moment of the electronic transition, decreasing the resonator mode volume and $\omega_c$, or enhancing the overlap of the photonic mode and the electronic polarization field. Following these considerations, the regime of ultrastrong coupling was first established using intersubband transitions of quantum wells (QWs) coupled to mid-infrared or THz waveguides [5-7,9] or plasmonic dot cavities [8,13], achieving values on the order of $\Omega_R/\omega_c \approx 0.25$ [13]. Circuit QED systems followed up shortly [10], with current values of up to $\Omega_R/\omega_c = 1.34$ [11,12]. In comparison, in the optical regime, the giant dipole moment of cyclotron resonances (CR) coupled to planar metamaterials has enabled a milestone value of $\Omega_R/\omega_c = 0.87$ [14,15], highlighting key advantages of the CR for QED [25,26] such as the tunability by the magnetic bias and control of the oscillator strength by the Landau level filling factor.

Here, we report an unprecedentedly large relative coupling strength of $\Omega_R/\omega_c = 1.43$ in planar THz metamaterials ultrastrongly coupled to CRs. In particular, we tailor the vacuum mode of the structure by both, the metamaterial and the electronic excitation. In conventional design approaches, where the electronic and photonic components are typically treated separately, the electronic resonance is placed in a maximum of the light field calculated for the empty resonator. For $\Omega_R/\omega_c$ exceeding unity, however, the electronic resonance strongly influences the light field, requiring a joint treatment of both components [19]. We have thus developed a parameter-free model of light-matter coupling which renders the electric field distribution on extremely sub-wavelength scales, enables predictive design of nanoscale QED structures, and an unambiguous identification of polariton modes. Exceeding unity relative coupling strength in the optical regime, for the first time, our structures enter uncharted terrain of non-perturbative light-matter dynamics, bringing theoretically discussed quantum phenomena expected for $\Omega_R/\omega_c > 1$ within experimental reach [16-20]. In particular, we give a quantitative experimental scenario for the most spectacular prediction, the detection of optical quantum vacuum radiation.

We employ high quality quantum wells hosting two-dimensional electron gases which are Landau-quantized by an external magnetic field of up to $B = 5$ T. The emerging CR with a frequency of $\nu_c = eB/2\pi m^*$, freely tunable by $B$, where $e$ is the elementary charge and $m^*$ is the effective mass, is coupled to the near-field of metamaterials processed on top of the structures. The transmission spectrum is measured using THz time-domain magneto-spectroscopy (Fig. 1a and Methods). Structural and optical parameters are summarized in Table 1.

We first investigate a conceptually straightforward structure consisting of a single QW with a doping density of $\rho = 3.0 \times 10^{11}$ cm$^{-2}$, located at a depth $z_0 = 500$ nm below the surface, and a double-gap metamaterial with a resonance frequency of $\nu_m = 0.82$ THz (Fig. 1a and Supplementary Information). The transmission spectra as a function of $\nu_c$ show the lower polariton (LP) which merges into the metamaterial resonance for large $\nu_c$, and the upper polariton (UP) (Fig. 1b, dashed lines). On resonance, $\nu_c = \nu_m$, the typical anti-crossing signature is observed and the spectral separation of LP and UP with frequencies of $\nu_{LP} = 0.74$ THz and $\nu_{UP} = 0.92$ THz, respectively, is minimal. In addition, the bare CR originating from uncoupled areas of the structure is visible as a diagonal line. Using a fitting procedure based on the Hopfield formalism which takes the entire polariton dispersion into account [14], we determine a coupling strength of $\Omega_R/\omega_c = 0.11$.

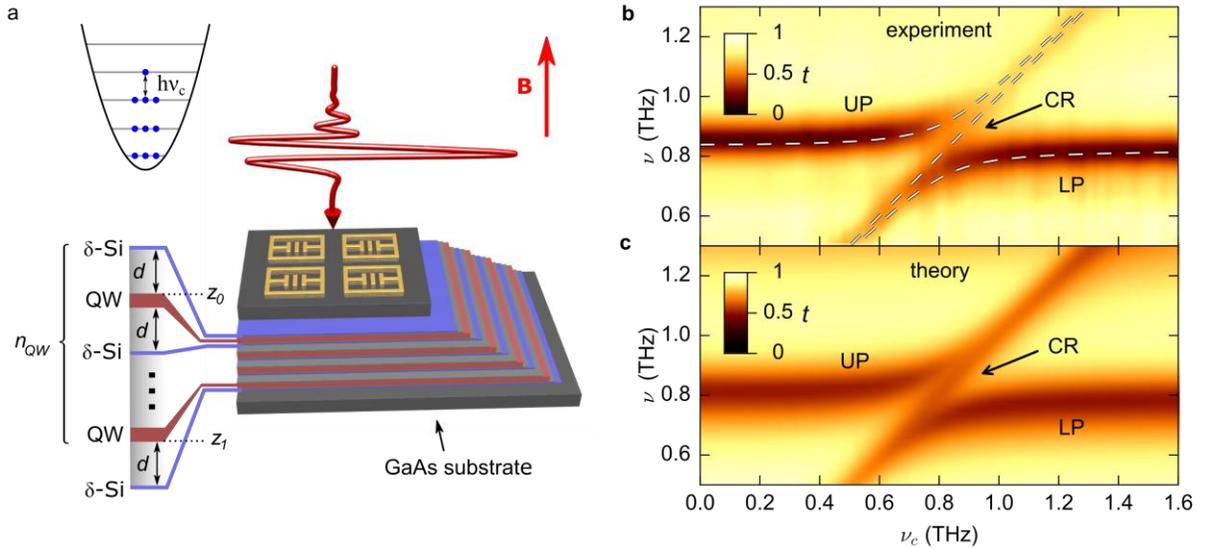

**Fig. 1 | Ultrastrong coupling of THz metamaterials to the cyclotron resonance of 2DEGs.** **a** Experimental geometry of magneto-THz transmission measurements. Red arrow: B-field. QW: quantum wells. δ-Si: doping layers. $z_0$, $z_1$: Upper and lower extension limit of the QW stack, respectively. Inset: Landau level fan of magnetically biased two-dimensional electron gas. **b** Color plot of measured transmission as a function of $\nu_c$. Dashed curves: Dispersion of the lower (LP) and upper polariton (UP) determined by fitting the transmission minima to the Hopfield Hamiltonian, which yields $\Omega_R/\omega_c = 0.11$. Diagonal line: cyclotron resonance. **c** Calculated transmission spectrum, and resonances labeled as in **b**. The Hopfield fit yields $\Omega_R/\omega_c = 0.10$.

We numerically calculate the complex field distribution and transmission spectra by a classical electrodynamical theory. Our formalism models the dielectric environment of the nanostructure and implements the CR as a gyrotropic medium, allowing for a unified treatment of photonic and electronic components and accounting for their mutual interplay (see Methods). The resulting far-field calculation (Fig. 1c) predicts the experimental transmission across the entire spectral range with high accuracy, and the Hopfield fit yields $\Omega_R/\omega_c = 0.10$.

As a first measure to increase $\Omega_R/\omega_c$, we exploit the relation $\Omega_R/\omega_c \propto (n_{QW} \times \rho)^{1/2}$ [27] and boost the oscillator strength of the Landau-quantized 2DEG by increasing $\rho$ to $5.4 \times 10^{11}$ cm$^{-2}$. We furthermore employ $n_{QW} = 25$ QWs of a thickness $d_{QW} = 25$ nm, each, separated from their doping layers by $d_\delta = 50$ nm to form a stack extending from $z_0 = 80$ nm to $z_1 = 3.1$ µm (cf. Fig. 1a). These modifications are accompanied by significantly more complex transmission spectra (Fig. 2a) which exhibit multiple resonances, requiring a strategy to unambiguously identify the polariton modes. Our theory matches the experimental data accurately (Fig. 2b), enabling this identification by two independent methods: First, we calculate the frequency evolution of each mode as a function of $\rho$, and verify that LP and UP merge into the metamaterial resonance for $\rho \to 0$, without any discontinuities. Second, we compare the spatial shapes of the polariton and metamaterial modes.

Following these procedures (see Supplementary Information), we pinpoint the LP and UP (dashed curves) with centre frequencies of $\nu_{LP} = 0.53$ THz and $\nu_{UP} = 1.38$ THz at the anti-crossing point, and $\Omega_R/\omega_c = 0.53$ as confirmed by our calculation, which yields $\Omega_R/\omega_c = 0.57$.

Analyzing the full spatial electric field distribution of the coupled system for this structure demonstrates that a separate treatment of resonator and electronic transition cannot accurately describe the system: The envelope function of the LP mode in z-direction (Fig. 2c, blue curve) significantly deviates from the mode of the bare resonator (black curve). In particular, the dense charge carrier plasma of the QWs (blue-shaded area) acts similarly to a mirror, leading to a localization of the electric field amplitude between the metamaterial and the QWs (magnified area), and an exponential decay inside the stack. This strong mutual interplay of the electronic transition and the resonator mode opens up new perspectives to control the shape of the vacuum mode by tailoring the electronic excitation. A particularly exciting option is to significantly increase the coupling strength by minimizing the mode volume not only by the design of the resonator, but by tuning the entire coupled system, including the heterostructure. The effectiveness of this approach is confirmed numerically in the more compact QW stack of structure 3 with the same metamaterial, where the localization of the mode is drastically enhanced, leading to an increased field amplitude inside the QW stack (Fig. 2d, red curve). Structure 3 is realized by reducing $d_\delta$ to 15 nm and $d_{QW}$ to 10 nm in a 6-fold QW stack, corresponding to a 14-fold reduction of the QW stack extension (Table 1) while $n_{QW} \times \rho$ is reduced by just 20% compared to structure 2 (see Supplementary Information). The spectra (Fig. 2e) remain qualitatively similar to structure 2 for both the experiment and the simulation. However, we determine $\nu_{LP} = 0.45$ THz and $\nu_{UP} = 1.47$ THz at the anti-crossing point, and the Hopfield fitting procedure yields an increased $\Omega_R/\omega_c = 0.62$ in very good agreement with our numerical calculation ($\Omega_R/\omega_c = 0.66$).

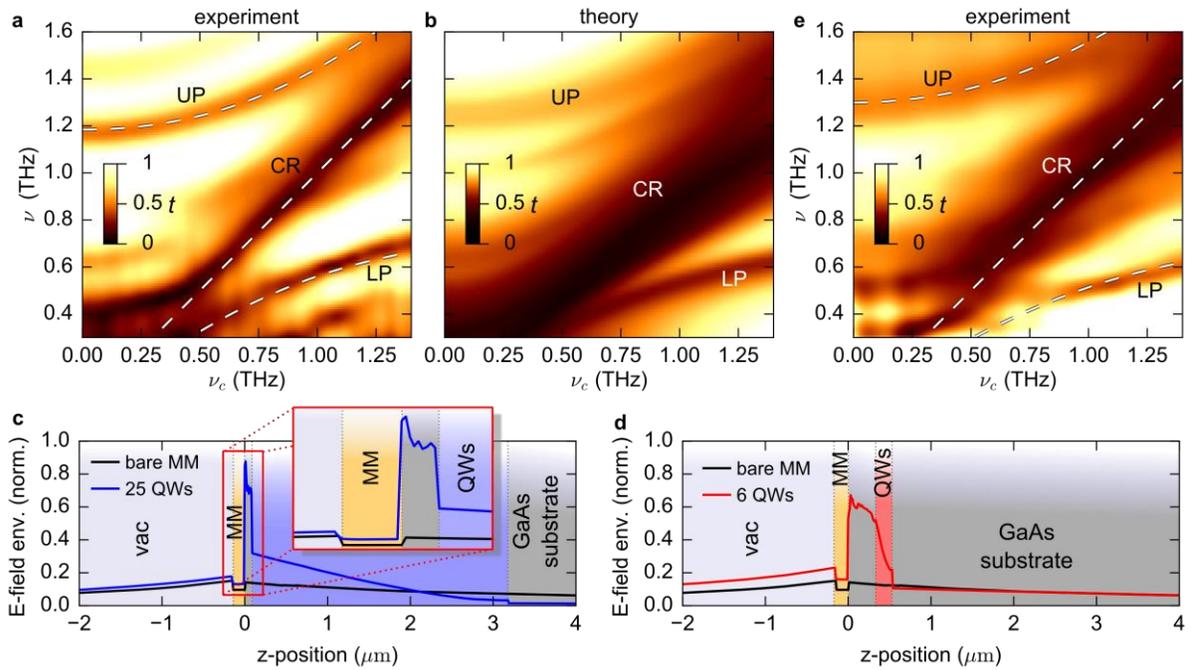

**Fig. 2 | Shaping of the vacuum mode by tailoring of the electronic excitation.**
**a** Color plot of the experimental transmission spectra of the 25-fold QW stack (structure 2). Dashed curves: polariton resonances obtained by Hopfield fit, which yields $\Omega_R/\omega_c = 0.53$. **b** Numerically calculated transmission (Hopfield fit: $\Omega_R/\omega_c = 0.57$). **c** Electric field envelope of the LP mode of structure 2, in z-direction (blue curve), averaged across the x-y-plane. The QW stack is shown as a blue-shaded area. The magnified area highlights the mode localization between the metamaterial, and the QWs. Black curve: Envelope of the bare metamaterial. **d** Significantly enhanced mode confinement of the 6-fold QW stack of structure 3 (red curve), and **e** resulting experimental transmission spectra, for which the Hopfield fit yields $\Omega_R/\omega_c = 0.62$.

Following this promising route of shaping the vacuum mode by the 2DEG, we push the limits of expitaxial heterostructure growth in a series of three 6-fold QW structures. We improve the mirror-like role of the 2DEG by simultaneously optimizing $d_\delta$ for more compact QW stacks and $z_0$ for higher mode localization. Exploiting the scaling $\Omega_R/\omega_c \propto \omega_c^{-1/2}$ [27] we employ a metamaterial with $\nu_m = 0.48$ THz implemented as an inverted structure following Babinet's principle [28], such that LP and UP manifest themselves as local transmission maxima.

| Structure | $n_{QW}$ | $\rho$ ($10^{12}$ cm$^{-2}$) | $z_0$ (nm) | $z_1$ (nm) | $d_\delta$ (nm) | $\Omega_R/\omega_c$ (exp.) | $\Omega_R/\omega_c$ (theory) |
|---|---|---|---|---|---|---|---|
| 1 | 1 | 0.30 | 500 | - | 30 | 0.11 | 0.10 |
| 2 | 25 | 0.54 | 80 | 3105 | 50 | 0.53 | 0.57 |
| 3 | 6 | 1.85 | 325 | 535 | 15 | 0.62 | 0.66 |
| 4 | 6 | 1.38 | 50 | 310 | 20 | 0.99 | 1.09 |
| 5 | 6 | 1.75 | 45 | 255 | 15 | 1.13 | 1.21 |
| 6 | 6 | 3.00 | 40 | 200 | 10 | 1.43 | 1.57 |

**Table 1 | Structural and optical parameters of the samples.**

For structure 4, the LP and UP resonances emerge as distinct features (Fig. 3a) separated by 0.95 THz at the anti-crossing point, corresponding to $\Omega_R/\omega_c = 0.99$ (theory: $\Omega_R/\omega_c = 1.09$, Fig. 3b). Intriguingly, light-matter coupling in these structures is strong enough to push a second-order lower polariton branch labelled LP2, resulting from strongly off-resonant coupling of the CR to a higher mode of the metamaterial (see Supplementary Information), into the spectral range of the fundamental LP and UP. Structures 5 and 6 exhibit qualitatively identical spectra (Figs. 3c,d) well-reproduced by our theory (Fig. 3e). Yet, the LP and UP resonances are pushed even further apart (Fig. 3f, top) resulting in coupling strengths of $\Omega_R/\omega_c = 1.13$ (theory: $\Omega_R/\omega_c = 1.21$) and $\Omega_R/\omega_c = 1.43$ (theory: $\Omega_R/\omega_c = 1.57$), respectively. These values mark the first realization of light-matter coupling strengths exceeding unity, in an optical system. Our optimization process demonstrates the power of the classical sub-wavelength calculations for boosting $\Omega_R/\omega_c$ by precisely predicting transmission spectra and revealing the role of tuning parameters such as $d_\delta$ (Supplementary Information), while allowing for unambiguous identification of spectral features.

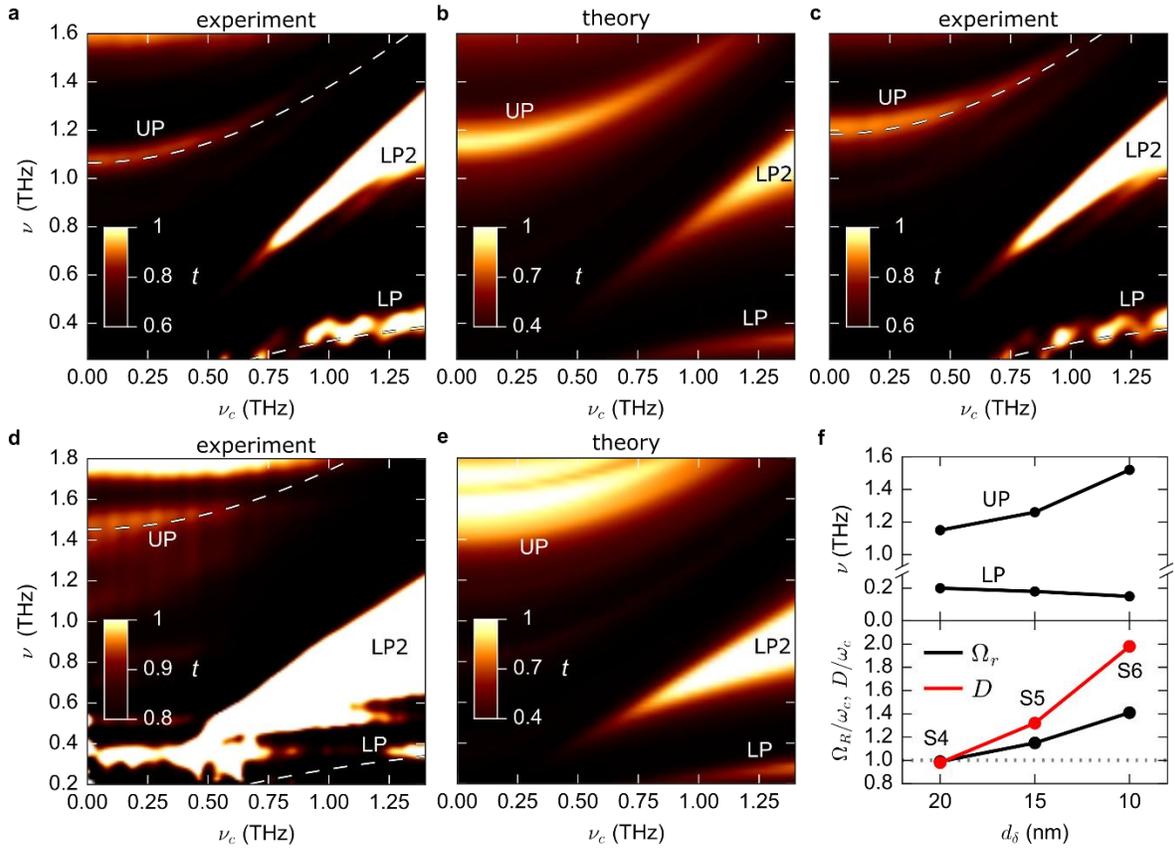

**Fig. 3 | Coupling of inverted metamaterials and CR of extremely compact 6-fold QW stacks. a** Transmission color plot of structure 4 with $d_\delta = 20$ nm and $\rho = 1.38 \times 10^{12}$ cm$^{-2}$. LP and UP are evident as local transmission maxima, and $\Omega_R/\omega_c = 0.99$. LP2: higher-order coupled mode (see text). **b** Corresponding calculation. **c** Structure 5; $d_\delta = 15$ nm, $\rho = 1.75 \times 10^{12}$ cm$^{-2}$, $\Omega_R/\omega_c = 1.13$. **d** Structure 6; $d_\delta = 10$ nm, $\rho = 3.00 \times 10^{12}$ cm$^{-2}$, $\Omega_R/\omega_c = 1.43$, and **e** calculation. **f** $\nu_{LP}$ and $\nu_{UP}$ for the 6-fold QW structures 4-6 (S4-S6) (top), and Rabi ($\Omega_R$) and diamagnetic coupling frequency ($D$) at the anti-crossing point (bottom).

At a relative coupling strength of $\Omega_R/\omega_c = 1.43$ we enter a completely new regime for optical systems where light-matter energy exchange occurs on a time scale significantly shorter than the lifetime of vacuum fluctuations as permitted by the energy-time uncertainty principle, boosting the virtual photon population of the ground state $|G\rangle$ [16]. Here, a formal treatment of the vacuum requires a full quantum theory, including anti-resonant light-matter interaction terms, $H_{anti} = i\hbar\Omega_{R,k}(\hat{a}_k\hat{b}_{-k} - \hat{a}_k^\dagger\hat{b}_{-k}^\dagger) + \hbar D_k(\hat{a}_k\hat{a}_{-k} - \hat{a}_k^\dagger\hat{a}_{-k}^\dagger)$, where $\hat{a}_k$ and $\hat{b}_k$ represent the annihilation operators for photon and matter excitations in mode $k$, respectively [16]. The latter term is linked to the virtual photon population of the ground state [16,20], and is proportional to the diamagnetic interaction strength $D_k \approx \Omega_{R,k}^2/\omega_c$ which scales quadratically with $\Omega_R$, on resonance. For structure 6, $D_k/2\pi = 0.98$ THz even exceeds the vacuum Rabi frequency $\Omega_R/2\pi = 0.69$ THz, i.e., anti-resonant exchange dominates over vacuum Rabi interactions (Fig. 3f) and a vacuum photon population of $\langle G|\hat{a}_k^\dagger\hat{a}_k|G\rangle = 0.37$ results. This novel scenario now enables experiments where diabatic switching [6,9] of $10^3$ resonators should lead to THz field amplitudes of 0.15 V/cm, at the detector (Supplementary Information). The incoherent nature of quantum vacuum radiation necessitates quantum detection [21,22] and statistical analysis, for which extremely low-noise electro-optic detection [29,30] enables single-shot sensitivities of 0.8 V/cm. Applying this detection scheme to our structures thus enables the clear isolation of quantum vacuum radiation from the noise floor within a few seconds of acquisition time (Supplementary Information).

Our approach introduces a paradigm change for the design of ultrastrongly coupled optical systems, opening up the uncharted terrain of coupling strengths beyond unity by treating the electronic and the photonic components of the system as a single entity. Our theory accounts for the complex interplay of the electronic polarization and the near-field distribution on sub-wavelength scales, enabling us to exploit the electronic excitation not only to boost the oscillator strength, but also to efficiently control the shape of the vacuum mode. Combining a predictive theory with advanced heterostructure growth provides a novel platform for the design of next-generation nanoscale QED systems. With up to $\Omega_R/\omega_c = 1.43$, our structures are a key step towards experimental quantum vacuum photonics, enabling the exploration of exotic quantum effects including a reversal of the Purcell effect, quantum vacuum phase transitions, and quantum vacuum radiation.

**Methods**

**Nanofabrication:**

Our semiconductor heterostructures were grown by molecular beam epitaxy on undoped (100)-oriented GaAs substrates which were prepared by growing an epitaxial GaAs layer of a thickness of 50 nm followed by an $Al_{0.3}Ga_{0.7}As$/GaAs superlattice to obtain a defect-free, atomically flat surface.

The GaAs quantum well (QW) stacks were embedded in $Al_{0.3}Ga_{0.7}As$ barriers with a total extension from a depth $z_0$ to $z_1$ below the sample surface. Si δ-doping layers were placed symmetrically around the individual QWs at a distance $d_\delta$ (Fig. 1a), enabling control of the carrier density $\rho$ of the two-dimensional electron gases (2DEGs) which is given as a density per QW throughout the manuscript, and of the electron mobility $\mu$ relevant for the Q-factor of the CR. The QW stacks were capped by a few-nm-thick GaAs layer for protection against oxidation. In order to tailor the electronic structure, the band profiles in growth direction were simulated using a Schrödinger-Poisson-formalism.

Metamaterial resonators forming the optical cavity were fabricated on the surface of the semiconductor structures by electron beam lithography of Polymethylmethacrylat (PMMA) resist followed by thermal vapour-phase deposition of 5 nm of Ti, improving adhesion of the subsequently deposited 150-nm-thick Au layer.

**Characterization:** Van der Pauw magneto-transport measurements of the two-dimensional electron gases were performed at a temperature of 4.2 K in a magnet cryostat in order to determine the electron mobilities and charge carrier densities of the two-dimensional electron gases. The effective electron masses $m^*$ of $0.070\,m_e$ (structure 1), $0.073\,m_e$ (structure 2), $0.080\,m_e$ (structures 3-6) were determined by terahertz cyclotron resonance spectroscopy.

**Terahertz magnetospectroscopy:**

Linearly polarized single-cycle terahertz pulses were generated in a ZnTe crystal by optical rectification of 35-fs pulses at a centre wavelength of 800 nm derived from a Ti:sapphire amplifier laser system. The pulses were normally incident on the sample structures which were cooled to 8 K in a magnet cryostat, and biased by a magnetic field of up to 5 T oriented perpendicularly to the sample surface. The transmitted pulses were electro-optically sampled in a second ZnTe crystal using 35-fs gate pulses from the amplifier.

**Numerical analysis:**

We numerically calculate the complex field distribution and transmission spectra of our nanostructures by a classical electrodynamical theory using finite-difference frequency-domain methods. For the metamaterial, we chose a dielectric constant of $\epsilon_{Au} = 10^5 + 10^5 i$ leading to a close to perfect metallic behaviour, while the undoped dielectric components were modelled using $\epsilon_{GaAs} = 12.9$. The CR of the Landau-quantized two-dimensional electron gas leads to a gyrotropic response which we implemented by introducing a dielectric tensor of a plasma of charge carriers magnetically biased along the z-direction [31],

$$\epsilon_{CR} = \begin{pmatrix} \epsilon_{xx} & i\epsilon_{xy} & 0 \\ -i\epsilon_{xy} & \epsilon_{xx} & 0 \\ 0 & 0 & \epsilon_{GaAs} \end{pmatrix},$$

where the components $\epsilon_{xx}(\omega) = \epsilon_{GaAs} - \dfrac{\omega_p^2(\omega+i\nu)}{\omega[(\omega+i\nu)^2 - \omega_c^2]}$ and $\epsilon_{xy}(\omega) = \dfrac{\omega_p^2 \omega_c}{\omega[(\omega+i\nu)^2 - \omega_c^2]}$ describe the two-dimensional polarization response of the CR in the plane perpendicular to the magnetic field.

The oscillator strength is given by the plasma frequency $\omega_p^2 = \rho e^2/\epsilon_0 m^*$, and the scattering rate $\nu = 3 \times 10^{11}$ Hz is chosen such that the experimentally determined linewidth is reproduced. Along the z-direction, the confinement of the QWs inhibits a plasma response such that we employ the background dielectric constant for GaAs, $\epsilon_{\text{GaAs}}$. The large span of geometrical scales in our system, ranging from the vacuum wavelength of the incident radiation exceeding 1 mm down to the thickness of the QWs on the order of 10 nm, implies a significant numerical complexity. In order to enhance numerical convergence, we thus apply a homogeneous dielectric function for the QW stack including the barrier material, with a uniform charge carrier density equivalent to the measured total electron sheet density of all quantum wells combined.

We numerically solve Maxwell's equations including linear light-matter interaction terms in the Hamiltonian, i.e., using the dipole approximation. Periodic boundary conditions are chosen to reflect the array character of our structure. We relate the calculated near-field profile to the far-field transmission $\vec{E}_f$ using the formalism of Stratton and Chu [32]:

$$\vec{E}_f(\omega) = \frac{ik}{4\pi} \hat{r}_0 \times \int_{\partial V} d^2r \left[\hat{n}(\vec{r}) \times \vec{E}(\vec{r}) - \epsilon(\vec{r})^{-1/2} \hat{r}_0 \times (\hat{n}(\vec{r}) \times \vec{H}(\vec{r}))\right] e^{ik\vec{r}\cdot\hat{r}_0},$$

where $k = \omega/c\sqrt{\epsilon(\vec{r})}$ is the wavenumber, $\hat{r}_0$ is the unit vector pointing in the far-field direction, $\hat{n}(\vec{r})$ is the normal vector of the surface, $\epsilon(\vec{r})$ is the local dielectric constant, and $\vec{E}(\vec{r})$ and $\vec{H}(\vec{r})$ are the electric and magnetic near-field components, respectively. The integration is performed across the surface $\partial V$ of the simulation volume $V$ enclosing the resonators, as a function of THz frequency $\nu$, and cyclotron frequency $\nu_c$.

**Acknowledgements**

We gratefully acknowledge support by the European Research Council through grant no. 305003 (QUANTUMsubCYCLE) and the Deutsche Forschungsgemeinschaft (LA 3307/1-1, HU 1598/2-1, BO 3140/3-1, and Collaborative Research Center SFB 689).

**Author contributions**

A.B., R.H., D.B., and C.L., conceived the study, A.B., M.P., S.S., and D.B. fabricated the samples, A.B., M.P., S.S., and C.L. performed the THz experiments, A.B., R.H., D.B. and C.L. wrote the manuscript. All authors discussed the results.

**Competing financial interests:** The authors declare no competing financial interests.